\documentstyle[prc,aps,preprint]{revtex}

\draft
\begin{document}
\tighten

\title{Exactly solvable model illustrating far-from-equilibrium
         predictions}
\author{O. Mazonka$^a$, C. Jarzynski$^b$}
\address{~ \\ (a) Institute for Nuclear Studies,
                \' Swierk, Poland \\
         {\tt maz@iriss.ipj.gov.pl}}
\address{(b) Theoretical Division, Los Alamos National
                Laboratory, USA \\
         {\tt chrisj@lanl.gov}}

\maketitle

\begin{abstract}
We describe an exactly solvable model which illustrates
the {\it fluctuation theorem} and other predictions for systems 
evolving far from equilibrium.
Our model describes a particle dragged by a spring
through a thermal environment.
The rate at which the spring is pulled is arbitrary.
\end{abstract}

\pacs{
      LAUR-99-6303
     }

\section*{Introduction}
\label{sec:intro}

In recent years, a number of theoretical results
pertaining to systems evolving far from thermal equilibrium
have been derived; see, for instance, 
Refs.\cite{es,gc,kurchan,crooks,ls,maes}.
While not identical, these results bear
a similar structure, and the term {\it fluctuation
theorem} has come to refer to them collectively.
Our purpose in this paper is to illustrate these 
and other\cite{dft,nwr} results, with a highly idealized 
but exactly solvable model of a system far from equilibrium.

The paper is organized as follows.
In Section \ref{sec:theory} we briefly review the
four theoretical predictions which we
will illustrate with our model.
These are:
the {\it steady-state} and {\it transient} fluctuation
theorems, a {\it detailed} fluctuation theorem,
and a {\it nonequilibrium work relation} for free
energy differences.
In Section \ref{sec:solve} we introduce and solve
our model, representing a particle dragged through
a thermal environment by a uniformly translating harmonic 
force.
In Section \ref{sec:illustrate}
we use this solution to show that our model indeed
satisfies the above-mentioned nonequilibrium results.

\section{Background}
\label{sec:theory}

Statements of the fluctuation theorem (FT) which 
have appeared in the literature can be expressed 
by the following equation:
\begin{equation}
\label{eq:ft}
\Bigl[
\lim_{\tau\rightarrow\infty}
\Bigr]
{1\over\tau}\ln
{p_\tau(+\overline\sigma)\over p_\tau(-\overline\sigma)}
= \overline\sigma.
\end{equation}
Here, $\overline\sigma$ is the average rate at which entropy
is generated during a given time interval 
-- or {\it segment} -- of duration $\tau$,
for a system away from thermal equilibrium;
$p_\tau(\overline\sigma)$ is the distribution of values of
$\overline\sigma$ over a statistical ensemble of such
segments; and Boltzmann's constant $k_B=1$.
(Note that the words ``average rate'' denote the time 
average over a given segment, {\it not} an average over 
the ensemble of segments.)
Following the literature, we will distinguish between
two versions of the FT,
{\it steady-state} and {\it transient},
which differ in the statistical 
ensemble of segments considered.  
In the steady-state case\cite{gc}, 
we imagine observing the system
in a nonequilibrium steady state for
an ``infinite'' length of time, which we chop up into
infinitely 
many segments of duration $\tau$, and we compute
the average entropy generation rate, $\overline\sigma$,
for each segment; $p_\tau(\overline\sigma)$
is then the distribution of values of $\overline\sigma$
over this ensemble of segments.
In the transient case\cite{es}, 
by contrast, we imagine that
the system of interest begins in an equilibrium state,
but is then driven {\it away} from
equilibrium, for instance by the sudden application
of an external force.
We observe the response of the system for a 
time $\tau$, starting from the moment the
external perturbation is applied. 
Then $p_\tau(\overline\sigma)$ is the distribution of
values of average entropy generation rate over infinitely 
many repetitions of this process.
The transient FT is valid for any duration $\tau$,
whereas the steady-state FT becomes valid
as $\tau\rightarrow\infty$;
hence the parenthetical appearance of that limit 
in Eq.\ref{eq:ft}.

In the examples considered in the literature,
the physical origin of entropy generation is the
exchange of heat between the system and some
infinite reservoir.
To model the evolution of a system in contact
with a reservoir, a variety of schemes are
available,
and in each case one must {\it define} what
is meant by the ``entropy generated''.
The FT has been derived or illustrated numerically
for a number of such schemes, involving both
deterministic and stochastic equations of motion.
Of particular relevance for the present paper
is the work of Kurchan\cite{kurchan}, who
showed that the FT is valid for Langevin processes.

Apart from the usual (steady-state and transient)
statements of the FT, 
a {\it detailed fluctuation theorem}
(DFT) was derived in Ref.\cite{dft}.
This result pertains to 
a finite time interval $\tau$, and has a
structure similar to that of the usual FT, but
in addition exhibits a dependence on the initial and
final microstates of the system.
Specifically, consider a process 
$\Pi^+$ during which a system
of interest evolves in contact with a heat
reservoir\footnote{
In fact, the formulation of Ref.\cite{dft} is
somewhat more general than that described here:
the system is allowed to be in contact with
numerous reservoirs, at various temperatures;
furthermore, contact between the system and any of the 
assorted reservoirs can be externally established and 
broken during the course of the process.
}, while -- possibly -- some {\it work
parameter} $\lambda$ (e.g.\ an applied field) is
being manipulated externally.
Let $\lambda^+(t)$ be the externally imposed 
time-dependence of this parameter, from $t=0$ to $t=\tau$,
and let $\Delta s$ denote the entropy produced during
a particular realization of this process.
Now consider the ``reverse'' process, $\Pi^-$,
defined exactly as $\Pi^+$, but with the time-dependence
of the work parameter reversed:
$\lambda^-(t)=\lambda^+(\tau-t)$.
For the forward process $\Pi^+$,
let $P_+({\bf z}_f,\Delta s\vert{\bf z}_i)$
denote the joint probability that the entropy produced
over the interval of observation will be $\Delta s$,
and the final microstate of the system will be
${\bf z}_f$, {\it given} an initial microstate
${\bf z}_i$;
and let $P_-({\bf z}_f,\Delta s\vert{\bf z}_i)$
denote the same for the reverse process, $\Pi^-$.
Then the DFT states that these joint, conditional
probability distributions satisfy the following
relation:
\begin{equation}
\label{eq:dft}
{P_+({\bf z}_B,+\Delta S\vert {\bf z}_A) \over
 P_-({\bf z}_A^*,-\Delta S\vert {\bf z}_B^*) } =
 \exp (\Delta S/k_B),
\end{equation}
where the asterisk ($^*$) denotes a reversal of momenta:
$({\bf q},{\bf p})^* = ({\bf q},-{\bf p})$.

Finally, in Ref.\cite{nwr} the following
{\it nonequilibrium work relation} for free energy
differences was established:
\begin{equation}
\label{eq:nwr}
\langle e^{-\beta W}\rangle =
e^{-\beta\Delta F}.
\end{equation}
(See also Ref.\cite{nwr_gc} for an elegant
derivation of
Eq.\ref{eq:nwr} specific to stochastic processes.)
This result applies to a situation in which a system of
interest, in contact -- and initially in equilibrium --
with a heat reservoir at temperature $\beta^{-1}$, 
evolves in time as an external parameter
$\lambda$ is switched {\it at a finite rate} from
and initial to a final value, say, from 0 to 1.
The finite-rate switching of $\lambda$ drives the
system out of equilibrium.
$W$ is the work performed during one realization
of this process.
The precise value of $W$ will depend on the microscopic
initial conditions of both the system of interest
and the reservoir;
by repeating the process infinitely many times 
(sampling initial
conditions from equilibrium ensembles), we obtain
a statistical ensemble of microscopic realizations
of the process.
The angular brackets then denote an average over
this ensemble of realizations, and 
$\Delta F$ is the free energy difference between 
the equilibrium states (at temperature $\beta^{-1}$) 
corresponding to the values $\lambda=0$ and 
$\lambda=1$.\footnote{
The validity of Eq.\ref{eq:nwr} does {\it not}
require that the system end in the equilibrium
state corresponding to $\lambda=1$,
only that such an equilibrium state exists,
and that the final value of the external 
parameter is 1.}
Eq.\ref{eq:nwr} is valid regardless of how slowly
or quickly $\lambda$ is switched from 0 to 1,
hence applies even if the system is driven far
from equilibrium by a violent variation of the
external parameter.

Eqs.\ref{eq:ft}, \ref{eq:dft}, and \ref{eq:nwr}
represent general theoretical predictions applicable
to systems far from thermal equilibrium.
We now introduce and solve a simple model which
illustrates these results.

\section{The model}
\label{sec:solve}

In this section we introduce our (one-dimensional) model
of a particle dragged through a thermal medium,
and then we solve it exactly:
given the particle's initial location, 
we compute the joint probability distribution of the final 
location and the net external work performed on the particle,
after an arbitrary time of evolution.
In the following section we
use this result to confirm both the steady-state
and transient fluctuation theorems,
as well as the detailed fluctuation theorem 
and the nonequilibrium work relation for free energy
differences.

In our model, the particle under consideration obeys
Langevin dynamics.
This model can therefore by viewed (with qualifications,
see Section \ref{subsec:eceg}) as a special case
of the situation considered by Kurchan\cite{kurchan},
who showed that the FT is satisfied for
this class of stochastic dynamics.
In essence, {\it we are illustrating Kurchan's results with a
specific model}.
However, the definition of entropy generated which 
we choose differs from that of Ref.\cite{kurchan}, 
and so we are obliged to express the steady-state
and transient FT's differently, in terms of
{\it power delivered} rather than {\it entropy
generation rate}.
We stress that this difference (between the formulation
of the FT given below, and that in Kurchan's work)
is only one of terminology, not substance.

Consider the following situation.
A particle, in contact with a thermal medium at temperature
$\beta^{-1}$, is pulled through that medium by
a time-dependent external harmonic force.
Assuming a single degree of freedom,
let $x$ denote the location of the particle, and let
\begin{equation}
\label{eq:pot}
U(x,t) = {k\over 2} (x-ut)^2
\end{equation}
be the moving potential well which drags the particle.
We can picture the particle as being attached to a
spring, the other end of which moves with a constant
speed $u$.
Assume furthermore that the thermal forces can be
modeled as the sum of linear friction and white noise,
and that the motion of the particle is overdamped.
Then the equation of motion for the position of the
particle is:
\begin{equation}
\label{eq:initial_eom}
\dot x = -{k\over\gamma} (x-ut) + \tilde v,
\end{equation}
where $\dot x \equiv dx/dt$, $\gamma$ is the coefficient of 
friction, and $\tilde v(t)$ represents delta-correlated 
white noise with variance $2/\beta\gamma$
(as mandated by the fluctuation-dissipation relation):
\begin{equation}
\langle\tilde v(t_1)\tilde v(t_2)\rangle = 
(2/\beta\gamma)\cdot\delta(t_2-t_1).
\end{equation}

Imagine that we observe the evolution of such a
particle over a time interval from $t=0$ to $t=\tau$,
and from the observed trajectory
we compute the total work $W$ performed by the
external potential over this interval.
Then the central result of this section will
be an answer to the following question.
{\bf Given an initial location $x_0$,
what is the joint probability distribution for the
final location and value of work performed $(x,W)$?}

To answer this question, we first introduce a
``work accumulated'' function, $w(t)$, which gives
the work performed on the particle up to time $t$;
hence, $W=w(\tau)$.
This function satisfies\cite{sekimoto}
\begin{equation}
\label{eq:work_eom}
\dot w(t) = {\partial U\over\partial t}\Bigl(x(t),t\Bigr)
= -uk \Bigl(x(t)-ut\Bigr),
\end{equation}
along with the initial condition $w(0)=0$.

It will furthermore prove convenient to specify the 
location of the particle by a variable $y=x-ut$ 
(i.e.\ in the reference frame of the moving well), 
rather than $x$.
Under this change of variables, Eqs.\ref{eq:initial_eom}
and \ref{eq:work_eom} become:
\begin{mathletters}
\label{eq:eom}
\begin{eqnarray}
\dot y &=& -{k\over\gamma} y - u + \tilde v \\
\dot w &=& -uky.
\end{eqnarray}
\end{mathletters}
Now imagine a statistical ensemble of such particles,
represented by an evolving probability distribution
$f(y,w,t)$. 
Eq.\ref{eq:eom} then translates into the Fokker-Planck
equation
\begin{equation}
\label{eq:fp}
{\partial f\over\partial t} =
{k\over\gamma}{\partial\over\partial y}(yf)
+ u {\partial f\over\partial y} +
uky {\partial f\over\partial w} +
{1\over\beta\gamma}
{\partial^2f\over\partial y^2}.
\end{equation}
What we now want is an expression for
$f(y,w,t\vert y_0)$, by which we mean the solution
to Eq.\ref{eq:fp} satisfying the initial conditions
\begin{equation}
f(y,w,0\vert y_0) = \delta(y-y_0) \delta (w).
\end{equation}
The function $f(y,w,t\vert y_0)$ is the joint probability
distribution for achieving a location $y$ and a value of
work accumulated $w$, at time $t$, given $y_0$ at time 0.
By evaluating this solution at $t=\tau$, and making
the change of variables from $y$ back to $x$, we
have the answer to the question posed earlier in
boldface.

To solve for $f(y,w,t\vert y_0)$, we first note that
Eq.\ref{eq:fp} has the following property: if at one 
instant in time the distribution happens to be Gaussian, 
then it will remain Gaussian for all subsequent times.
This follows from the fact that the drift and diffusion
coefficients in Eq.\ref{eq:fp} are either constant or
linear in $y$ and $w$.
Now, a normalized Gaussian distribution $f^G(y,w)$ is 
uniquely defined by the values of the following moments:
\begin{mathletters}
\label{eq:define_moments}
\begin{eqnarray}
\hat y &\equiv& \int f^G(y,w) \, y \\
\hat w &\equiv& \int f^G(y,w) \, w \\
\sigma_y^2 &\equiv& \int f^G(y,w) \, (y^2 - \hat y^2) \\
\sigma_w^2 &\equiv& \int f^G(y,w) \, (w^2 - \hat w^2) \\
c_{yw} &\equiv& \int f^G(y,w) \, (yw - \hat y\hat w),
\end{eqnarray}
\end{mathletters}
where the integrals are over $(y,w)$-space.
An explicit expression for $f^G(y,w)$ in terms of 
these moments is:
\begin{mathletters}
\label{eq:gaussian_moments}
\begin{equation}
f^G(y,w) = {\sqrt{C}\over 2\pi}\,\exp(-{\bf z}^T{\bf C}{\bf z}/2),
\end{equation}
where 
\begin{equation}
{\bf z} =
\left(
\begin{array}{c}
y-\hat y \\ \\ w-\hat w
\end{array}
\right)
\qquad , \qquad
{\bf C} =
\left(
\begin{array}{cc}
C\sigma_w^2 & -Cc_{yw}     \\ \\
-Cc_{yw}    & C\sigma_y^2
\end{array}
\right),
\end{equation}
\end{mathletters}
$C = (\sigma_y^2 \sigma_w^2 - c_{yw}^2)^{-1} = \det {\bf C}$,
and ${\bf z}^T$ denotes the transpose of ${\bf z}$.

The evolution of a time-dependent Gaussian distribution
$f^G(y,w,t)$ is thus uniquely specified by the evolution
of the moments $\hat y,\cdots,c_{yw}$.
Given a distribution evolving under Eq.\ref{eq:fp},
we get from Eq.\ref{eq:define_moments}
the following set of coupled equations of motion 
for these moments:
\begin{mathletters}
\label{eq:eom_moments}
\begin{eqnarray}
\label{eq:single_moments}
{d\over dt} \hat y = -{k\over\gamma} \hat y - u
\qquad &,& \qquad
{d\over dt} \hat w = -uk \hat y \\
\label{eq:double_moments}
{d\over dt} \sigma_y^2 = -{2k\over\gamma} \Bigl(\sigma_y^2 -
{1\over\beta k}\Bigr) 
\quad,\quad
{d\over dt} c_{yw} &=& - {k\over\gamma} (c_{yw} + u\gamma \sigma_y^2)
\quad,\quad
{d\over dt} \sigma_w^2 = -2uk c_{yw}.
\end{eqnarray}
\end{mathletters}

The distribution $f(y,w,t\vert y_0)$ will thus be a Gaussian
whose moments evolve according to Eq.\ref{eq:eom_moments}, 
and satisfy the initial conditions:
$\hat y=y_0$,
$\hat w = \sigma_y^2 = c_{yw} = \sigma_w^2 = 0$.
The solution, as easily verified by substitution, is: 
\begin{mathletters}
\label{eq:solmom}
\begin{eqnarray}
\hat y(t\vert y_0) &=& -l + (y_0+l) e^{-kt/\gamma} \\
\label{eq:sol_y}
\hat w(t\vert y_0) &=& uklt + u\gamma(y_0+l) (e^{-kt/\gamma}-1) \\
\sigma_y^2(t\vert y_0) &=& {1\over\beta k}(1-e^{-2kt/\gamma}) \\
c_{yw}(t\vert y_0) &=& -{u\gamma\over\beta k} (e^{-kt/\gamma}-1)^2 \\
\label{eq:sol_sy}
\sigma_w^2(t\vert y_0) &=& {u^2\gamma^2\over\beta k}
\Bigl(
{2kt\over\gamma} - e^{-2kt/\gamma} + 4 e^{-kt/\gamma} - 3
\Bigr),
\end{eqnarray}
\end{mathletters}
where
\begin{equation}
l={\gamma u\over k}.
\end{equation}

The combination of Eqs.\ref{eq:solmom} and 
\ref{eq:gaussian_moments} gives the solution for
$f(y,w,t\vert y_0)$ for all times $t\ge 0$
(and for all values of $y$, $w$, and $y_0$).
The explicit expression given in the Appendix.
Note that, by projecting out either of the two independent
variables $y$ and $w$, we can obtain the marginal probability 
distributions for the other:
\begin{mathletters}
\begin{eqnarray}
\rho(y,t\vert y_0) &\equiv&
\int dw\,f(y,w,t\vert y_0) =
{1\over\sqrt{2\pi\sigma_y^2}}
\exp[-(y-\hat y)^2/2\sigma_y^2] \\
\label{eq:eta}
\eta(w,t\vert y_0) &\equiv&
\int dy\,f(y,w,t\vert y_0) =
{1\over\sqrt{2\pi\sigma_w^2}}
\exp[-(w-\hat w)^2/2\sigma_w^2].
\end{eqnarray}
\end{mathletters}

It is instructive to consider the limit of
asymptotically long times, $kt/\gamma\gg 1$.
In this limit,
\begin{mathletters}
\label{eq:asymp}
\begin{eqnarray}
\label{eq:asymp_y}
\hat y(t\vert y_0) &\rightarrow& -l \\
\label{eq:asymp_w}
\hat w(t\vert y_0) &\rightarrow& uklt + {\it O}(1) \\
\label{eq:asymp_sy}
\sigma_y^2(t\vert y_0) &\rightarrow& 1/\beta k \\
c_{yw}(t\vert y_0) &\rightarrow& -u\gamma/\beta k \\
\sigma_w^2(t\vert y_0) &\rightarrow& 2u^2\gamma t/\beta + {\it O}(1),
\end{eqnarray}
\end{mathletters}
where the ``order unity'' corrections to
$\hat w(t\vert y_0)$ and $\sigma_w^2(t\vert y_0)$ 
represent terms which
converge to a constant as $t\rightarrow\infty$.
We see that, for any $y_0$, 
the distribution of positions settles into a 
steady-state Gaussian of variance $1/\beta k$,
centered at a displacement $-l$
from the minimum of the confining potential $U$
(Eqs.\ref{eq:asymp_y}, \ref{eq:asymp_sy}).
Note that a {\it canonical} (equilibrium) distribution of 
positions would have the same variance, but centered at the 
minimum.
Hence, $l$ represents the extent to which the steady-state 
distribution of positions ``lags behind'' the instantaneous
equilibrium distribution.
This lag could have been predicted with an educated guess,
by ignoring fluctuations and simply balancing the
frictional and harmonic forces, $-\gamma u$ and $-ky$.
One can similarly understand the leading behavior of
$\hat w(t)$: if the position of the particle is (on average) 
a distance $l$ behind the instantaneous minimum of the
well, then the harmonic spring is pulling the particle
with a force $+kl$, at a velocity $u$, hence delivering 
a power $ukl$.

\subsection{Energy conservation and entropy generation}
\label{subsec:eceg}

In our model, energy conservation translates into
the following balance equation
between the external work $W$ performed
on the particle, the heat $Q$ absorbed from the thermal
surroundings, and the net change in the internal energy
of the particle\cite{sekimoto}:
\begin{equation}
\label{eq:firstlaw}
W + Q = \Delta U,
\end{equation}
where 
$\Delta U = U\Bigl(x(\tau),\tau\Bigr)-U\Bigl(x(0),0\Bigr)$,
$W=w(\tau)$, and
\begin{equation}
Q = \int_0^\tau \dot x(t)\,
{\partial U\over\partial x}
\Bigl(x(t),t\Bigr)\,dt.
\end{equation}

In addition to these quantities ($W$, $Q$, and $\Delta U$),
we need a microscopic definition of the
{\it entropy generated} over one realization of the
process.
There is necessarily some arbitrariness in such a definition,
but we will use the following one:
\begin{equation}
\label{eq:entropy_definition}
\Delta S \equiv -\beta Q.
\end{equation}
Thus, we identify the entropy generated with the amount
of heat dumped into the environment ($-Q$), divided
by the temperature.
This definition  -- motivated by macroscopic thermodynamics
(see e.g.\ Ref.\cite{dft}) --
differs from that of Kurchan\cite{kurchan},
who identifies the entropy production rate with the
external power delivered to the system, divided by the
reservoir temperature.
Both definitions seem to be reasonable, but we will
use Eq.\ref{eq:entropy_definition}.

Quite apart from the definition of entropy generation,
our set-up differs from Kurchan's in the way in which
external work is defined;
see Section 2.2 of Ref.\cite{kurchan}.
Nevertheless, the general analytical approach taken by
Kurchan applies to our set-up as well.
Hence, any results derived in Ref.\cite{kurchan}
regarding average power delivery (external work performed,
divided by the duration of the time interval), 
extend to our situation.

\section{Illustration of far-from-equilibrium predictions}
\label{sec:illustrate}

We now use the results of the previous section
to show that Eqs.\ref{eq:ft}-\ref{eq:nwr} 
are obeyed by our model.

\subsection{Transient and steady-state fluctuation theorems}
\label{subsec:tssft}

We begin with the transient and the
steady-state fluctuation theorems.
Ordinarily the FT is expressed in terms of the average 
entropy production rate (Eq.\ref{eq:ft}).
By contrast, the relations which we will show to be satisfied 
by our model (Eq.\ref{eq:wft}) are expressed in 
terms of the average {\it power delivered} to our particle
as it is dragged through its thermal environment.
This difference -- as mentioned earlier -- is a consequence 
of our choice of definition of average entropy generation
rate, $\overline\sigma = \Delta S/\tau = -\beta Q/\tau$.
(Kurchan, by contrast, defines entropy generation
in terms of power delivered,
$\overline\sigma_{\rm Kurchan}=\beta W/\tau$.\cite{kurchan})
In the calculations to follow, the reader should bear
in mind that the transient and steady-state
relations which we establish {\it are essentially
those of Kurchan};
only our definition of entropy production
prevents us from presenting them as such.

(It might be interesting to investigate whether
or not Eq.\ref{eq:ft} remains valid under our
definition of entropy production,
$\overline\sigma = -\beta Q/\tau$.
This would be easy to check with numerical
simulations, but in the ``exactly solvable''
spirit of restricting ourselves to analytical
results, we have not pursued this question.)

For a time interval of duration $\tau$, let
\begin{equation}
X = W/\tau
\end{equation}
denote the time-averaged rate at which work is performed on
the particle -- i.e.\ the average power delivered -- by the
moving harmonic potential.
Now imagine that we observe the particle for an ``infinitely''
long time as it evolves in the steady state;
we divide this time of observation into infinitely many
segments of duration $\tau$; we compute the average
power delivered, $X$, over each segment; and we construct
the statistical distribution of these values,
$p_\tau^S(X)$.
We will obtain an explicit expression for $p_\tau^S(X)$ for
our model, and will show that it satisfies the following
{\it steady-state} FT:
\begin{mathletters}
\label{eq:wft}
\begin{equation}
\label{eq:sswft}
\lim_{\tau\rightarrow\infty} {1\over\tau}
\ln {p_\tau^S(+X)\over p_\tau^S(-X)}
=\beta X.
\end{equation}

Now imagine instead that we begin with the particle in thermal 
equilibrium, and then we drag it for a time $\tau$ with the
harmonic confining potential.
(That is, for $t<0$ the potential is stationary,
with the particle in equilibrium;
then, between $t=0$ and $t=\tau$ the potential is
moved rightward with velocity $u$.)
Again defining $X$ to be the average power delivered over
the interval $0\le t\le\tau$, 
let $p_\tau^C(X)$ denote the statistical
distribution of values of $X$, over infinitely many repetitions
of this process, always starting from equilibrium.
\footnote{
The superscript $C$ indicates that the particle's
initial conditions are sampled from a canonical ensemble.}
We will solve for $p_\tau^C(X)$ and show that it obeys the
following {\it transient} FT:
\begin{equation}
\label{eq:twft}
{1\over\tau}\ln {p_\tau^C(+X)\over p_\tau^C(-X)}
=\beta X,
\end{equation}
\end{mathletters}
whose validity does {\it not} require the limit
$\tau\rightarrow \infty$.

Let us begin by considering the steady-state case.
Since the right sides of Eqs.\ref{eq:asymp_y} and
\ref{eq:asymp_sy} are independent of $y_0$,
an arbitrary initial distribution of particle
positions $\rho(y,0)$ will settle into a Gaussian:
\begin{equation}
\label{eq:ss}
\lim_{t\rightarrow\infty}
\rho(y,t) = 
\sqrt{\beta k\over 2\pi}
\exp[-\beta k(y+l)^2/2]
\equiv \rho^S(y),
\end{equation}
which defines the instantaneous distribution of
particle positions in the steady state.
Then $p_\tau^S(X)$ -- the distribution of values
of average power delivered, $X=W/\tau$, over time
intervals of duration $\tau$ sampled during the
steady state -- can be constructed
by folding together $\rho^S(y_0)$ and 
$\eta(w,\tau\vert y_0)$ (Eq.\ref{eq:eta}):
\begin{equation}
p_\tau^S(X) = 
\int dw\,
\delta(X-w/\tau)\,
\int dy_0\,
\rho^S(y_0)\,
\eta(w,\tau\vert y_0).
\end{equation}
Here, the integral over $dy_0$ produces the 
distribution of values of work, after time $\tau$, 
given initial conditions sampled from the steady
state.
The integral over $dw$ converts that distribution 
of values of $w$ into one of values of $X$.
Examining Eqs.\ref{eq:sol_y}, \ref{eq:sol_sy},
\ref{eq:eta}, and \ref{eq:ss},
we see that, in the product 
$\rho^S(y_0)\eta(w,\tau\vert y_0)$,
the variable $y_0$ appears only in powers up
to the quadratic inside an exponent;
hence this product is a Gaussian in $y_0$,
and the integral can be carried out explicitly.
Without going through the (modestly tedious)
details, we present the result:
\begin{equation}
p_\tau^S(X) =
{1\over\sqrt{2\pi\sigma_X^2}}
\exp\Biggl[-{(X-\overline X)^2\over 2\sigma_X^2}\Biggr],
\end{equation}
where
\begin{equation}
\label{eq:Xsigmu}
\overline X = ukl \quad,\quad
\sigma_X^2 = {2\mu\overline X\over\beta\tau} \quad,\quad
\mu(\tau) = 1 + {\gamma\over k\tau}
(e^{-k\tau/\gamma}-1).
\end{equation}
($\overline X$ represents the average instantaneous
power delivered to the particle, in the steady state;
see the comments following Eq.\ref{eq:asymp}.)
From this we obtain an explicit expression for the
left side of Eq.\ref{eq:sswft}:
\begin{equation}
{1\over\tau}
\ln {p_\tau^S(+X)\over p_\tau^S(-X)}
= {1\over\tau}\cdot
{2\overline X X\over\sigma_X^2} =
{\beta X\over\mu}.
\end{equation}
Since $\lim_{\tau\rightarrow\infty}\mu=1$,
we conclude that the FT
for power delivered in the steady state (Eq.\ref{eq:sswft})
is indeed satisfied for our model.
Note, however, that the limit $\tau\rightarrow\infty$
{\it is} necessary.

In the case of the {\it transient} FT,
pertaining to a system driven away from an initial
state of canonical equilibrium, we can
construct $p_\tau^C(X)$ in the same way as
$p_\tau^S(X)$, only now we fold $\eta$ in with
a canonical distribution,
$\rho^C(y)\propto\exp(-\beta ky^2/2)$,
rather than the steady-state distribution:
\begin{eqnarray}
\label{eq:pc}
p_\tau^C(X) &=&
\int dw\,
\delta(X-w/\tau)\,
\int dy_0\,
\rho^C(y_0)\,
\eta(w,\tau\vert y_0) \\
&=&
{1\over\sqrt{2\pi\sigma_X^2}}
\exp\Biggl[-{(X-\mu\overline X)^2\over 2\sigma_X^2}\Biggr],
\end{eqnarray}
where $\sigma_X^2$, $\overline X$, and $\mu(\tau)$ are
exactly as above.
The only difference between the steady-state and the
transient distribution of values of $X$ is, we see,
the factor $\mu$ appearing inside the exponent
in the latter.
This small difference has the effect that the
FT for power delivered in the transient case
is satisfied for all positive values of $\tau$:
\begin{equation}
{1\over\tau}
\ln {p_\tau^C(+X)\over p_\tau^C(-X)}
= {1\over\tau}\cdot
{2\mu\overline X X\over\sigma_X^2} =
\beta X.
\end{equation}

\subsection{Detailed fluctuation theorem}
\label{subsec:dft}

We now show that our model satisfies
the detailed fluctuation theorem (DFT).
As mentioned in Section \ref{sec:theory}, 
the DFT is stated in terms of two
processes, $\Pi^+$ (``forward'') and $\Pi^-$ (``reverse''),
related by time-reversal.
In the present context, we take $\Pi^+$ to be the
process studied above:
a particle is dragged through a thermal medium
by a time-dependent potential well,
\begin{equation}
U^+(x,t) = {k\over 2} (x-ut)^2.
\end{equation}
The reverse process, $\Pi^-$, is then obtained by
moving the well in the opposite direction:
\begin{equation}
U^-(x,t) = {k\over 2} (x+ut-u\tau)^2.
\end{equation}
Formally, we can think of the minimum of the potential
as being given by $\lambda \Delta x$, where 
$\Delta x = u\tau$ is a constant, and $\lambda$
is an externally controlled parameter.
During the process $\Pi^+$, $\lambda$ is changed
uniformly from 0 to 1;
during $\Pi^-$, from 1 to 0.

Again taking $y$ to be the displacement of the particle
relative to the instantaneous minimum of the potential, 
we define $f_+(y,w,t\vert y_0)$ and 
$f_-(y,w,t\vert y_0)$ to be the joint probability
distributions for attaining $(y,w)$ at time $t$,
given $y_0$ at time 0, for the two processes.
We then solve for these two distributions exactly
as we solved for $f(y,w,t\vert y_0)$ in
Section \ref{sec:solve}.
The solutions are:
\begin{eqnarray}
f_+(y,w,t\vert y_0) &=&
f(y,w,t\vert y_0) \\
f_-(y,w,t\vert y_0) &=&
f(y,w,t\vert y_0)_{u\rightarrow -u},
\end{eqnarray}
where $f$ is just the solution obtained
in Sec.\ref{sec:solve}.
Thus, the solution for the forward process
is identical to the solution of Sec.\ref{sec:solve}
(as it must be, since $\Pi^+$ is exactly the process
studied there!),
whereas the solution for the reverse process is
obtained from $f$ by replacing $u$
by $-u$ everywhere, {\it including in the definition
of $l$} (hence, $l\rightarrow -l$).
This replacement is easily understood:
in terms of the variable $y$, the process $\Pi^-$
is no different than that obtained by starting
with the minimum of the potential at $x=0$, and 
moving it with velocity $-u$ for a time $\tau$.

Let us now put these solutions to good use.

The DFT, in the context of this problem, claims the 
following:
\begin{equation}
\label{eq:dft1}
{P_+(y_B,+\Delta S\vert y_A) \over
 P_-(y_A,-\Delta S\vert y_B) } = \exp\Delta S,
\end{equation}
where $P_\pm(y_f,\Delta s\vert y_i)$ denote the
joint probability distributions of finding
the particle at a final point $y_f$, and a value
of entropy generated $\Delta s$, given an initial
location $y_i$, for the forward and 
reverse processes.\footnote{
Strictly speaking, these should be defined in terms
of absolute locations $x_A$ and $x_B$, but the change of
variables to relative displacements $y_A$ and $y_B$ is
immediate.}
Now consider a single realization of either process. 
Energy conservation, combined with our definition of entropy
produced (Eqs.\ref{eq:firstlaw} and \ref{eq:entropy_definition}),
give us the following relation between $y_i$, $y_f$, $\Delta s$,
and the work $w$ performed:
\begin{equation}
w - \beta^{-1}\Delta s =
{k\over 2} (y_f^2-y_i^2) \equiv
\Delta U(y_i,y_f).
\end{equation}
We can then re-express $P_\pm$ in terms of $f_\pm$:
\begin{eqnarray}
P_\pm(y_f,\Delta s\vert y_i) &=&
\int dw f_\pm(y_f,w,\tau\vert y_i)\,
\delta(\Delta s + \beta\Delta U - \beta w)\\
&=&
\beta^{-1} f_\pm(y_f,{k\over 2}(y_f^2-y_i^2)+
\beta^{-1}\Delta s,\tau\vert y_i),
\end{eqnarray}
from which we obtain the following explicit expressions
for the numerator and denominator in Eq.\ref{eq:dft1}:
\begin{eqnarray}
P_+(y_B,+\Delta S\vert y_A) &=& \beta^{-1}
f_+(y_B,+\Omega,\tau\vert y_A) \\
P_-(y_A,-\Delta S\vert y_B) &=& \beta^{-1}
f_-(y_A,-\Omega,\tau\vert y_B),
\end{eqnarray}
where
\begin{equation}
\Omega\equiv {k\over 2}(y_B^2-y_A^2) + 
\beta^{-1} \Delta S.
\end{equation}
Eq.\ref{eq:dft1} is thus equivalent to the
following relation:
\begin{equation}
\label{eq:dft_equiv}
{f_+(y_B,+\Omega,\tau\vert y_A) \over
 f_-(y_A,-\Omega,\tau\vert y_B)} = 
\exp\Bigl[
\beta\Omega - {\beta k\over 2}(y_B^2-y_A^2)
\Bigr],
\end{equation}
where, if Eq.\ref{eq:dft1} is to be valid for all
real $\Delta S$, then Eq.\ref{eq:dft_equiv} 
must hold for all real $\Omega$.
Using the expression for $f$ given
in the Appendix, one can verify that,
indeed, Eq.\ref{eq:dft_equiv} is valid for
all values of $\Omega$, hence
the DFT is satisfied by our model.

\subsection{Generalization}

Before proceeding to the nonequilibrium work relation, 
we point out that our
model is easily generalized to include an additional
uniform, constant external force.
Namely, suppose we modify the time-dependent potentials
$U^\pm(x,t)$ for the forward and reverse processes,
as follows:
\begin{equation}
\label{eq:generalization}
U^\pm(x,t) \rightarrow
U^\pm(x,t) + \alpha x \qquad,\qquad \alpha>0.
\end{equation}
This corresponds to subjecting the particle to an
additional leftward-pushing force, of magnitude
$\alpha$.
Thus, assuming $u>0$, the particle is dragged
``up'' the potential energy slope $\alpha x$
during $\Pi^+$, and ``down'' the slope during
$\Pi^-$.

The solution in this case, for the forward process
$\Pi^+$, is the same as that in the previous section,
except that $l$ is everywhere replaced by
\begin{equation}
l_\alpha = l + {\alpha\over k}
= {\alpha+\gamma u\over k}.
\end{equation}
Thus, in the steady state, the average position of
the particle is displaced by an amount $\alpha/k$
to the left, relative to the case with no additional
force (Eq.\ref{eq:asymp_y}).
This implies that additional work is performed
on the particle at an average rate
$u\alpha$ (Eq.\ref{eq:asymp_w});
this is simply the average rate at which we
drag the particle up the slope.

For the reverse process, the solution is obtained
(as in the previous Section) by the further
replacement $u\rightarrow -u$, including in the
definition of $l_\alpha$.

It is straightforward to show that, with these
replacements, the DFT remains valid.

\subsection{Nonequilibrium work relation for free 
energy differences}
\label{subsec:nwr}

Let us finally use the results derived above to show 
explicitly that Eq.\ref{eq:nwr} is satisfied by our model.
We will consider the process defined by $U^+(x,t)$ in the 
previous section (Eq.\ref{eq:generalization}).
As before, let $\Delta x=u\tau$ be a fixed distance,
and let $\lambda\Delta x$ define the minimum of
the confining potential, so that we move that minimum
from 0 to $\Delta x$ by changing $\lambda$ from 0 to 1:
\begin{eqnarray}
U_\lambda(x) &=& {k\over 2} (x-\lambda\Delta x)^2 + \alpha x \\
U^+(x,t) &=& U_{\lambda(t)}(x),
\end{eqnarray}
where $\lambda(t)=t/\tau$.
For any fixed value of $\lambda$ and fixed temperature
$\beta^{-1}$, there exists an equilibrium state of the
system, defined microscopically by a canonical
distribution.
The associated free energy $F_\lambda$ is then
defined in terms of the logarithm of the corresponding
partition function:
\begin{eqnarray}
F_\lambda &=& -\beta^{-1} \ln \int dx\,
e^{-\beta U_\lambda(x)} \\
&=& \alpha\lambda\Delta x - {\alpha^2\over 2k} +
{1\over 2\beta}\ln{\beta k\over 2\pi}.
\end{eqnarray}
Hence the free energy difference is simply
\begin{equation}
\label{eq:deltaf}
\Delta F = F_1 - F_0 = \alpha\Delta x.
\end{equation}
Physically, this is the work required to {\it reversibly}
(i.e.\ infinitely slowly)
change $\lambda$ from 0 to 1, at constant temperature.

Now consider a statistical ensemble of realizations
of our {\it finite-time},
irreversible process, $\lambda(t)=t/\tau$, 
with initial conditions
sampled from the canonical ensemble corresponding
to $\lambda=0$.
The associated distribution of values
of work, $W=w(\tau)$, can be written as:
\begin{equation}
\eta(W) = \int dy_0\,
\rho_\alpha^C(y_0)\,
\eta(W,\tau\vert y_0), \\
\end{equation}
where $\rho_\alpha^C(y_0)$ is the canonical distribution
of initial conditions (for $\alpha\ne 0$), and 
$\eta(W,\tau\vert y_0)$ is the distribution of
work values at time $\tau$, given initial condition
$y_0$.
We can use the results of Section \ref{sec:solve},
with the substitutions $l\rightarrow l_\alpha$
and $u=\Delta x/\tau$, to compute this integral explicitly.
Once again skipping the algebra, we present the
result:
$\eta(W)$ is a Gaussian distribution of
mean and variance 
\begin{eqnarray}
\label{eq:explicitW}
\langle W\rangle &=&
\alpha\Delta x + 
\mu\gamma (\Delta x)^2/\tau \\
\sigma_W^2 &=&
2 \mu\gamma (\Delta x)^2/\beta\tau,
\end{eqnarray}
with $\mu(\tau)$ as defined by Eq.\ref{eq:Xsigmu}.
(Note that in the reversible limit --
i.e.\ $\tau\rightarrow\infty$ with $\Delta x$ fixed --
we get $\langle W\rangle=\alpha\Delta x$
and $\sigma_W^2=0$, hence $W=\Delta F$ for every
realization, in agreement with the remarks following
Eq.\ref{eq:deltaf}.)
Thus, for any positive value of $\tau$, 
$\eta(W)$ {\it is a Gaussian whose mean and variance
are related by}
\begin{equation}
\label{eq:pseudofdt}
\langle W\rangle = 
\Delta F + \beta\sigma_W^2/2.
\end{equation}
This implies --
as can be verified by direct evaluation of
$\int dW \eta(W)\exp(-\beta W)$ --
that the nonequilibrium
work relation for free energy differences, 
Eq.\ref{eq:nwr} above, is satisfied.

Eq.\ref{eq:pseudofdt} has the structure
of the usual
{\it fluctuation-dissipation theorem} (FDT)
for linear response,
but differs from the latter in the fact that
the fluctuations -- represented by $\sigma_W^2$ --
are {\it not} obtained from the equilibrium
fluctuations of the particle,
but rather truly reflect behavior far from
equilibrium.
Indeed, the presence of $\mu(\tau)$ in 
Eq.\ref{eq:explicitW} is evidence of non-linear
response in our model: for fixed $\Delta x$, 
the average dissipated work, 
$\langle W\rangle - \Delta F$,
is not simply inversely proportional to $\tau$,
as would be the case for linear response.

\section*{Acknowledgments}

The authors gratefully acknowledge useful correspondence
with J. Kurchan, clarifying the important issue
of the definition of entropy generation.
This work was carried out during reciprocal visits to
Warsaw and Los Alamos, made possible by financial
support from the
Polish-American Maria Sk\l odowska-Curie Joint
Fund II, under project PAA / DOE-98-343,
as well as by the hospitality of the Institute
for Nuclear Studies (Poland) and Los Alamos National
Laboratory (USA).

\section*{Appendix}

Here we give an explicit expression for the function
$f(y,w,t\vert y_0)$ introduced in Section \ref{sec:solve}.
This solution is essentially the combination of 
Eqs.\ref{eq:gaussian_moments} and \ref{eq:solmom}:
\begin{equation}
f(y,w,t|y_0) = {\sqrt{C} \over 2\pi} \exp(AC/2),
\end{equation}
where
\begin{eqnarray}
C &=& \det{\bf C} = - { \beta^2 k^2 \over 2\gamma u^2 \nu_-[2\gamma \nu_-
+kt\nu_+] }\\
A &=& (1/ \beta k) \Bigl\{\nu_+\nu_-(w-kltu)^2 
+\gamma^2 u^2 \nu_-
(y-y_0) [ 4 l \nu_-+(3\nu-1)y_0
+(\nu-3)y ] \\
&&\quad +2\gamma u \Bigl( l^2 k u t \nu_-^2 -l\nu_-[2w\nu_-+
ktu\nu_+(y_0-y)]-
w \nu_-^2 (y_0+y)-
ktu(y-\nu y_0)^2 \Bigr)
\Bigr\},
\end{eqnarray}
and
\begin{equation}
\nu = \exp ( -k t / \gamma ) \quad,\quad
\nu_\pm = \nu \pm 1.
\end{equation}

\end{document}